\documentstyle[12pt,psfig]{article}
\begin{document}
\begin{center}
{ \large \bf ORIGINS OF THE BARYON SPECTRUM }
\end{center}

\medskip
\begin{center}
L. Ya. Glozman
\end{center}
\begin{center}
{\it High Energy Accelerator 
Research Organization (KEK),
Tanashi Branch, Tanashi, Tokyo 188-8501, Tokyo, Japan\\
 Institute for Theoretical Physics, 
        University of Graz, Universit\"atsplatz 5, A-8010 Graz,  Austria}
\end{center}


\begin{abstract}
I begin with  a key problem of
light and strange baryon spectroscopy which suggests
a clue for our understanding of underlying dynamics. Then I discuss
 spontaneous breaking of chiral symmetry in QCD,
which implies that at low momenta there must be quasiparticles -
constituent quarks with dynamical mass,  which should be coupled
to other quasiparticles - Goldstone bosons. Then it is natural
to assume that in the low-energy regime the underlying dynamics
in baryons is due to Goldstone boson exchange (GBE) between constituent
quarks. Using as a prototype of the microscopical
quark-gluon degrees of freedom the instanton-induced 't Hooft
interaction  I show why the GBE is so important. 
When the 't Hooft interaction is iterated in the qq t-channel
it inevitably leads to a pole which corresponds to GBE. 
This is a
typical antiscreening behavior: the interaction is represented by a bare
vertex at large momenta, but it blows up at small momenta in
the channel with GBE quantum numbers,
explaining thus a distinguished role of the latter interaction
in the low-energy regime. I show how the explicitly
flavour-dependent short-range part of the GBE
interaction between quarks, perhaps in combination with the
vector-meson exchange interaction, solves a key problem of baryon
spectroscopy and present spectra obtained in a simple analytical
calculation as well as in exact semirelativistic three-body
approach. 
\end{abstract}

\section{Where is a key problem?}

If one considers a model with an effective confining interaction
between quarks in light and strange baryons, which is flavour- and 
spin - independent\footnote{The Thomas precession, which is a
kinematical effect, and which produces a strong spin-orbit
force,  certainly presents in heavy quark systems, where
the heavy quark  constantly sits on the end of the string.
A relativistic rotation of the string implies the Thomas
precession. In the light quark systems, where it costs no
energy to break a string and the light quark  permanently
fluctuates into other quark and the quark-antiquark pair, this
kinematical effect should be  strongly suppressed.
That is why there are no strong spin-orbit splittings in light baryon
and meson spectra.} and assuming that there are no
residual interactions, then the spectrum of lowest baryons
should be arranged into successive bands of positive and
negative parity, see Fig. 1. In Nature, however, the lowest levels in
the spectra of nucleon, $\Delta$-resonance and $\Lambda$-hyperon,
which are  shown on Fig. 2, look pretty different.
 One can immediately conclude that
a  picture, where all other possible interactions
are treated as only residual and weak is certainly wrong.

\begin{figure}
\psfig{file=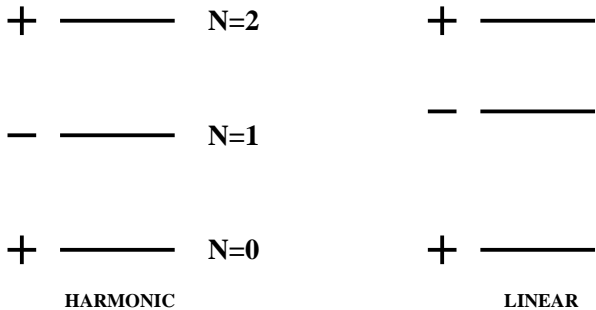}
\caption{A sequence of positive and negative parity 
levels with harmonic and linear confining interactions.}
\end{figure}

Typically models pay an attention to  the octet-decuplet
splittings. Within a quark picture one needs a spin-spin
force between valence quarks with a proper sign. Then, 
adjusting a strength of this spin-spin force one can explain
why $\Delta$ is heavier than nucleon, or why $\Sigma$ is
heavier than $\Lambda$ \cite{SAH}. When QCD appeared,  it has been immediately
suggested that such a spin-spin force is supplied by
the colour-magnetic component of the one gluon exchange (OGE)
\cite{RGG,MIT,IK}, in analogy with the magnetic hyperfine
interaction from the one photon exchange in quantum electrodynamics. 
At the price of a very large strong coupling constant, $\alpha_s \sim
1$, one can then fit $\Delta-N$ mass difference. Clearly that
such a picture is self-contradictory, because a big value of
$\alpha_s$ is not compatible with the perturbative treatment of QCD.

The crucial point, however, is that the perturbative gluon exchange
(does not matter, one gluon exchange or one thousand gluon
exchange) is sensitive only to spin (and colour) degrees of
freedom of quarks and there is no sensitivity at the operator
level to the flavour of quarks (in the u,d,s quark sector there
is only a very weak sensitivity via  different masses of
quarks  which, however, completely vanishes in the chiral
limit). The spin structure of all baryons in $N$ and $\Lambda$
spectra, depicted in Fig. 2, is the same, it is described by
the mixed  permutational symmetry. This means that the
contribution of the colour-magnetic interaction to leading
order is the same in all these baryons (up to some small
difference in baryon orbital wave functions), which is in 
apparent conflict with the {\it opposite} orderings of the lowest 
levels   in $N$ and
$\Lambda$ spectra. The only difference between $N$ and
$\Lambda$ system is that one light quark is substituted by
a strange one. It immediately hints that the physics,
responsible for Fig. 2, should be explicitly flavour dependent.
In addition, a colour magnetic interaction cannot shift
the $N = 2$ states $N(1440)$ and $\Lambda(1600)$ below
the $N=1$ states $N(1535)-N(1520)$ and 
$\Lambda(1670)-\Lambda(1690)$, respectively, because to
leading order its contribution is the same in all these
states. In the $\Delta$ spectrum the situation is even
more dramatic as the colour - magnetic interaction
shifts the $N=2$ state $\Delta(1600)$ up, but not down, 
with respect to the $N=1$ pair $\Delta(1620) - \Delta(1700)$. 

These facts rule out perturbative gluon exchange picture
as a source of the hyperfine interactions in the light and
strange baryons. 

\begin{figure}
\psfig{file=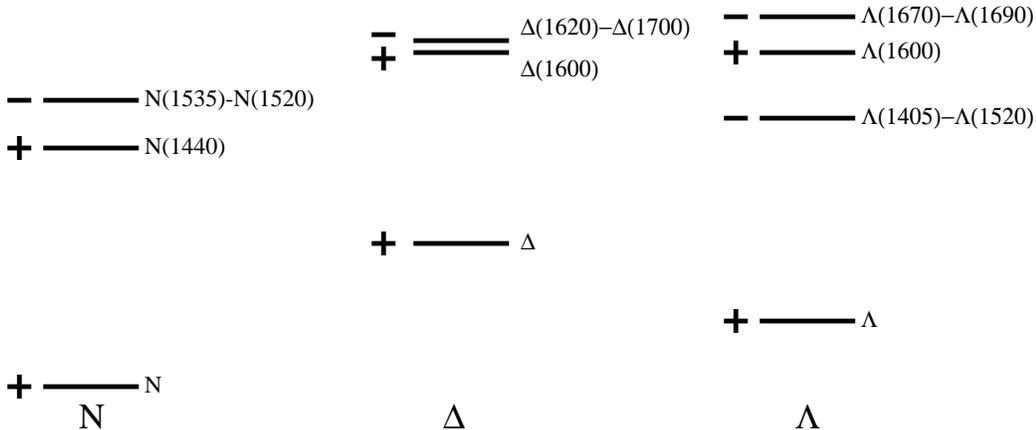}
\caption{Low-lying spectra of nucleon, $\Delta$-resonance and 
$\Lambda$-hyperon.}
\end{figure}

The other possible source of the hyperfine
interactions, the 't Hooft instanton induced interaction
\cite{HOOFT} between valence quarks, could, generally speaking, 
generate the octet-decuplet splittings \cite{KOCH,SR,BLASK}
when its strength is adjusted.
However, it is easy to see from its
operator structure that it also fails to explain Fig. 2 as
far as $N$ and $\Lambda$ parts are concerned. But the most
convincing evidence comes from the $\Delta$ spectrum, where
the 't Hooft interaction between valence quarks is identically
zero (it is absent in flavour - symmetric states). So according
to this scenario the $\Delta$ spectrum should be exclusively 
due to confining interaction, which is ruled out by
comparison of Figs. 1 and 2.  

Thus a key problem  is to explain {\it at the same time} both
the octet-decuplet splittings and the pattern of Fig. 2.

\section{Spontaneous breaking of chiral symmetry and its implications}

The  
$SU(3)_{\rm L} \times SU(3)_{\rm R}$ chiral symmetry of QCD Lagrangian
is spontaneously
broken down to $SU(3)_{\rm V}$ by the QCD vacuum (in the large $N_c$ limit
it would be $U(3)_{\rm L} \times U(3)_{\rm R} \rightarrow U(3)_{\rm V}$).
There are two important generic consequences of the spontaneous breaking
of chiral symmetry (SBCS). The first one is an appearance of the octet
of pseudoscalar mesons of low mass, $\pi, {\rm K}, \eta$, which represent
the associated approximate Goldstone bosons (in the large $N_c$ limit the
flavor singlet state $\eta'$ should be added). The second one is that valence
(practically massless) quarks acquire a dynamical mass, which has been called 
historically  constituent mass. Indeed, 
the nonzero value of the quark condensate, 
$<\bar q q> \sim -(250 MeV)^{3}$, 
itself implies at the formal level that there must be at low momenta
rather big dynamical mass, which should be  a momentum-dependent quantity. 
Such a dynamical mass is now directly observed on the lattice \cite{AOKI}.
Thus the constituent quarks should be considered as quasiparticles
whose dynamical mass at low momenta comes from the nonperturbative 
gluon and quark-antiquark dressing.
The flavour-octet axial current conservation in the chiral limit tells that
the constituent quarks and Goldstone bosons should be coupled with the
strength $g=g_A M/f_\pi$ \cite{WEIN}, which is a quark analog of the famous 
Goldberger-Treiman relation. 

We have recently suggested that in the low-energy regime,
below the chiral symmetry breaking scale, $\Lambda_\chi \sim 1$ GeV,
the low-lying light and strange baryons should be predominantly
viewed as systems of 3 constituent quarks with an effective
confining interaction and a chiral interaction mediated by a
Goldstone boson exchange (GBE) between the constituent quarks \cite{GR}.

\section{Why the Goldstone boson exchange is so important?}

Consider as example of a microscopical QCD nonperturbative
interaction the instanton-induced 't Hooft interaction
for two light flavours (I consider for simplicity a chiral limit)

\begin{equation}
H=-G[(\bar{\psi}\psi)^2 +(\bar{\psi}i \gamma_5 \vec{\tau} \psi)^2
-(\bar{\psi} \vec{\tau} \psi)^2 - (\bar{\psi}i \gamma_5  \psi)^2].
\label{HOOFT}
\end{equation}

\noindent
This interaction is known to lead to  chiral symmetry breaking,
i.e. to creation of the quark condensate and dynamical (constituent)
mass $m$ of quarks. It happens because of the first term in (\ref{HOOFT}),
which represents a scalar part of the interaction. The interquark
interaction in the pseudoscalar-isovector $\bar q q$ systems is
driven by the second term, which is attractive and so strong that
when it is iterated it exactly compensates the $2m$ energy
supplied by the first term, and thus there appear $T=1, J^P = 0^-$
mesons with zero mass - Nambu-Goldstone bosons. The first two
terms in the Hamiltonian above form  a classical Nambu and
Jona-Lasinio model \cite{NJL}. The fourth term in (\ref{HOOFT}),
which is repulsive, contributes only in the flavour-singlet
$\bar q q$ pair ($\eta'$),
making this meson heavy - contrary to $\pi$ - and solving
thus the $U(1)_A$ problem (note that the perturbative gluon exchange
force cannot solve it). There is no interaction term which
can contribute in vector mesons. This means that the masses
of vector mesons, $\rho$ and $\omega$,  
should be approximately $2m$, which is well satisfied empirically.
The interaction (\ref{HOOFT}), extended to the $u,d,s$ sector,
 also naturally explains completely
different mixing between the octet and singlet components in the
pseudoscalar and vector mesons \cite{HOOFT2}.

\begin{figure}
\psfig{file=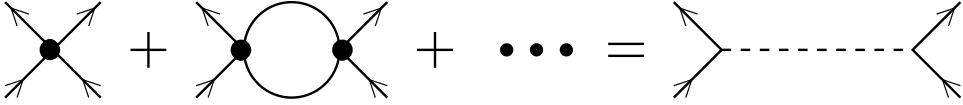}
\caption{Iteration of the instanton-induced 't Hooft interaction
in the t-channel. Black filled circle means a bare gluonic vertex.}
\end{figure}

Having mentioned all the positive features of the Hamiltonian
(\ref{HOOFT}) in mesons, I shall now discuss its implications
in baryons \cite{GV}. As I said, a direct application of this
instanton-induced interaction between valence quarks in baryons
does not solve problems. But what happens when this interaction
is iterated in $qq$ t-channel, see Fig. 3 ? Specifically, the
second term in (\ref{HOOFT}) will imply the following amplitude

\begin{equation}
T_P(q^2) = 2G + 2G J_P(q^2) 2G +... = 
\frac{2G}{1-2G J_P(q^2)},
\label{ITERP}
\end{equation}

\noindent
where $J_P(q^2)$ is a bubble  with a pseudoscalar vertex
(vacuum polarization in the pseudoscalar channel).
The denominator in (\ref{ITERP}) has a pole in the chiral
limit at $q^2=0$, which can be identified as a pion-exchange
(beyond the chiral limit it is shifted
to a physical pion mass $q^2=\mu_\pi^2$.) The coupling constant
of pion to constituent quark can be obtained as a residue
of (\ref{ITERP}) at the pole. The eq. (\ref{ITERP}) defines
a ``running amplitude'' and a negative sign in the denominator
implies its antiscreening behavior. In essence this antiscreening
is some kind of asymptotic freedom: at sufficiently large
space-like momenta the interaction is represented by a pure
't Hooft vertex (i.e. it has a strength $2G$), but at
$q^2 \rightarrow 0$ it becomes {\it infinitely} enhanced in
the channel with GBE quantum numbers.
So, if a typical momentum transfer is not large, which is the case
in baryons in the low-energy regime, the pole contribution
dominates. It explains why the GBE 
is so crucially important both in baryons and
baryon-baryon systems. Thus the GBE
interaction between constituent quarks is  an effective
representation of the pole contribution in (\ref{ITERP}),
which is provided by the original quark-gluon degrees of
freedom.

In fact any pairwise gluonic interaction between quarks in the
local approximation will necessarily contain the first and
second terms of (\ref{HOOFT}) with fixed relative strength.
This is because of chiral invariance. Thus all our conclusions
on $\pi - \rho$ mass splitting and Goldstone boson exchange
interaction in baryons are rather general and do not rely
necessarily on 't Hooft interaction.

\section{The Goldstone boson exchange interaction}

The
coupling of the constituent  quarks and the pseudoscalar Goldstone
bosons will (in the $SU(3)_{\rm F}$ symmetric approximation) have
the form $g/(2m)\bar\psi\gamma_\mu
\gamma_5\vec\lambda^{\rm F}
\cdot \psi \partial^\mu\vec\phi$ within the nonlinear realization of chiral
symmetry (it would be $ig\bar\psi\gamma_5\vec\lambda^{\rm F}
\cdot \vec\phi\psi$ within the linear  chiral symmetry 
representation). A coupling of this
form, in a nonrelativistic reduction for the constituent quark spinors,
will -- to lowest order -- give rise the 
$\sim\vec\sigma \cdot \vec q \vec\lambda^{\rm F}$ structure of the 
meson-quark vertex, where $\vec q$ is meson momentum. This type
of vertex implies spin-spin and tensor interactions between
constituent quarks, mediated by Goldstone bosons. 
The spin-spin force has a traditional
long-range Yukawa part, which is important for nuclear
force. But at short range the spin-spin force is much stronger
and its sign  is
{\it opposite}. This short-range interaction has a form \cite{GR}

\begin{equation} H_\chi\sim -\sum_{i<j}
\frac{V(\vec r_{ij})}{m_i m_j}
\vec \lambda^{\rm F}_i \cdot \vec \lambda^{\rm F}_j\,
\vec
\sigma_i \cdot \vec \sigma_j,\label{GBE} \end{equation}

\noindent
where a radial behavior of this short-range interaction is
unknown.
{\it It is this short-range part of the 
GBE interaction between the constituent quarks that is of crucial importance
for baryons: it has a sign appropriate to reproduce the level splittings
and strongly dominates over the Yukawa tail towards short distances.}
Note that this spin-spin force is explicitly flavour-dependent, which
reflects the fact that the GBE interaction
is a flavour-exchange one. It is also significant that this
short-range part of the interaction appears at the leading order
within the chiral perturbation theory (i.e. in the chiral limit)
\cite{GLOZ}, while the Yukawa part of the interaction vanishes
in this limit. This simple observation has by far-going
consequences: while the physics of baryons does not change
much in the chiral limit (e.g. the $\Delta - N$ mass splitting
persists), the long-range nuclear spin-spin force vanishes.
This means that in some sense the short-range part of the
pion exchange interaction is ``more fundamental'' than its
Yukawa part.

\section{The vector- and scalar-exchange interactions}

Already in ref. \cite{GR} it has been pointed out that the vector-like
meson exchange interactions could be also important. This
possibility is taken seriously in refs. \cite{G,WGPV}.
Both the vector- and scalar-meson exchange interaction can
be also considered as a representation of the correlated
two GBE interaction \cite{RB} as it
has a vector meson pole in t-channel. A phenomenological
motivation to include these interactions in addition to
one GBE is as follows. The spin-spin
component of the vector-meson exchange interaction at
short range has exactly the same flavor-spin structure (\ref{GBE}) as
one GBE, but their tensor force components
are  just of opposite sign and cancel each other to a big extent. 
This could explain an empirical
fact that the tensor force component of the interaction between
quarks in baryons should not be large. Otherwise it would cause small,
but empirically counterindicated spin-orbit splittings in
L=1 baryons. The small net tensor force component should be,
however, important for the mixing in baryon wave functions,
while the baryon mass is weakly sensitive to this small
residual tensor force. The present uncertainties in the coupling
constants and unknown short-range behavior of these effective
interactions make it very difficult to determine a precise
amount (and even sign) of this weak net tensor force from the
low-lying baryon spectroscopy. Other datum, e.g. mixing angles
extracted from strong and electromagnetic decays should be used
to determine the precise relative contributions of the effective
ps- and vector-exchanges.

The scalar- and vector-meson exchanges have  spin-orbit force
components. These spin-orbit forces are known to be very important
in $NN$ system, where both $\rho$- and $\omega$-exchange provide
spin-orbit force with the same sign in  P-wave. In baryons the
relative sign of these spin-orbit components becomes opposite in
P-wave (because of additional colour degree of freedom) and 
the $\rho$-exchange spin-orbit force becomes strongly enhanced
\cite{G}. This explains a weak net spin-orbit force in baryons,
while it is big and empirically very important in baryon-baryon
systems.

\section{The flavour-spin hyperfine interaction and the
structure of the baryon spectrum}

 Summarizing previous sections one concludes that the pseudoscalar-
and vector-meson exchange interactions produce strong 
flavour-spin interaction (\ref{GBE}) at short range while the net
tensor and spin-orbit forces are rather weak. That the net spin-orbit
and tensor interactions between constituent quarks {\it in baryons}
should be weak also follows from the typically small splittings
in LS-multiplets, which are of the order 10-30 MeV. These small splittings
should be compared with the hyperfine splittings produced by spin-spin
force, which are of the order of $\Delta - N$ splitting. Thus, indeed,
in baryons it is the spin-spin interaction  (\ref{GBE}) between constituent
quarks that is of crucial importance.

Consider first, for the purposes of illustration, a schematic model
which neglects the radial dependence
of the potential function $V(r)$ in (\ref{GBE}), and assume a harmonic
confinement among quarks as well as $m_{\rm u}=m_{\rm d}=m_{\rm s}$.
In this model
\begin{equation}H_\chi = -\sum_{i<j}C_\chi~
\vec \lambda^{\rm F}_i \cdot \vec \lambda^{\rm F}_j\,
\vec
\sigma_i \cdot \vec \sigma_j.\label{4} \end{equation}

The Hamiltonian (\ref{4})
 reduces the $SU(6)_{\rm FS}$  symmetry down to
 $SU(3)_{\rm F}\times SU(2)_{\rm S}$.
Let us now see how the pure confinement spectrum of Fig. 1 
becomes modified when
the  Hamiltonian (\ref{4}) is switched on.
The leading $SU(6)$ wave
functions are known for all low-lying baryons and we thus can
evaluate analytically  the expectation values of the
operator (\ref{4}) \cite{GR}.

For the octet states ${\rm N}$, $\Lambda$, $\Sigma$,
$\Xi$ ($N=0$ shell) as well as for their first
radial excitations of positive parity
 ${\rm N}(1440)$, $\Lambda(1600)$, $\Sigma(1660)$,
$\Xi(?)$ ($N=2$ shell) 
 the expectation value of the
Hamiltonian (\ref{4})
 is $-14C_\chi$. For the decuplet states
$\Delta$, $\Sigma(1385)$, $\Xi(1530)$, $\Omega$ ($N=0$ shell)
and their lowest radial excitations of positive parity $\Delta(1600)$
($N=2$) the corresponding matrix element is
$-4C_\chi$. In the  negative parity excitations
($N=1$ shell) in the ${\rm N}$, $\Lambda$ and $\Sigma$ spectra 
(${\rm N}(1535)$ - ${\rm N}(1520)$, $\Lambda(1670)$ - $\Lambda(1690)$
and $\Sigma(1750)$ - $\Sigma(?)$)
the contribution of the interaction (\ref{4})  is $-2C_\chi$. 
The first negative parity excitations in the $\Delta$ spectrum
$\Delta(1620)$ and $\Delta(1700)$ ($N=1$) produce the matrix element $4C_\chi$.
The first negative
parity excitation in the $\Lambda$ spectrum ($N=1$ shell)
$\Lambda(1405)$ - $\Lambda(1520)$ is flavor singlet 
and, in this case, the corresponding matrix element is $-8C_\chi$. The latter
state is unique and is absent in other spectra due to its flavour-singlet
nature.

These  matrix elements alone suffice to prove that
the ordering of the lowest positive and negative parity states
in the baryon spectrum will be correctly predicted by
the chiral boson exchange interaction (\ref{4}).
The constant $C_\chi$ may be extracted from the
N$-\Delta$ splitting to be 29.3 MeV.
The oscillator
parameter $\hbar\omega$, which characterizes the
effective confining interaction with this schematic model,
may be determined as  one half of the mass differences between the
first excited
$\frac{1}{2}^+$ states and the ground states of the baryons,
which have the same flavour-spin, flavour and spin symmetries
(e.g. ${\rm N}(1440)$ - ${\rm N}$, $\Lambda(1600)$ - $\Lambda$, $\Sigma(1660)$
- $\Sigma$),
to be
$\hbar\omega \simeq 250$ MeV. Thus the two free parameters of this simple model
are fixed and we can  make now predictions.

In the ${\rm N}$, $\Lambda$  and $\Sigma$ sectors the mass
difference between the lowest
excited ${1\over 2}^+$ states (${\rm N}(1440)$, $\Lambda(1600)$, 
and $\Sigma(1660)$)
and the ${1\over 2}^--{3\over 2}^-$ negative parity pairs
 (${\rm N}(1535)$ - ${\rm N}(1520)$, $\Lambda(1670)$ - $\Lambda(1690)$,
 and $\Sigma(1750)$ - $\Sigma(?)$, respectively) will then
be
\begin{equation}{\rm N},\Lambda,\Sigma:
\quad m({1\over 2}^+)-m({1\over 2}^--{3\over
2}^-)=250\, {\rm
MeV}-C_\chi(14-2)=-102\, {\rm MeV},\end{equation}
whereas for the lowest state ${1\over 2}^+$  in the $\Lambda$ 
system, $\Lambda(1600)$, and the lowest negative parity pair 
${1\over 2}^--{3\over 2}^-$
($\Lambda(1405)$ - $\Lambda(1520)$) it should be

\begin{equation}\Lambda:\quad m({1\over 2}^+)-m({1\over 2}^--{3\over
2}^-)=250\, {\rm
MeV}-C_\chi(14-8)=74\, {\rm MeV}.  \end{equation}

\noindent
At last, the lowest positive parity state ${3\over 2}^+$ in the $\Delta$
spectrum $\Delta(1600)$ \footnote{Note that the experimental
uncertainties for the mass of this baryon are 1550 - 1700 MeV.}
 should be approximately degenerate
with the lowest negative parity ${1\over 2}^--{3\over 2}^-$ 
excitations $\Delta(1620) -
\Delta(1700)$

\begin{equation}\Delta:\quad m({3\over 2}^+)-m({1\over 2}^--{3\over
2}^-)=250\, {\rm
MeV}-C_\chi(4+4)=15\, {\rm MeV}.  \end{equation}

One recovers  precisely the spectrum shown in Fig. 2. It
is astonishing that such a crude model predicts not only
a general structure of the low-lying spectrum, but also
the absolute values for splittings.

This simple example shows how the chiral interaction 
provides different ordering of the lowest positive and negative parity excited
states in the spectra of the nucleon and
the $\Lambda$-hyperon. This is a direct
consequence of the symmetry properties of the boson-exchange interaction
\cite{GR}.
Namely, completely symmetric FS state in the ${\rm N}(1440)$,
$\Lambda(1600)$ and
$\Sigma(1660)$ positive parity resonances from the $N=2$ band feels a
much stronger
attractive interaction than the mixed symmetry FS state  in the
${\rm N}(1535)$ - ${\rm N}(1520)$, $\Lambda(1670)$ - $\Lambda(1690)$
and $\Sigma(1750)$ -$\Sigma(?)$ resonances of negative parity ($N=1$ shell).
Consequently the masses of the
positive parity states ${\rm N}(1440)$, $\Lambda(1600)$  and
$\Sigma(1660)$ are shifted
down relative to the other ones, which explains the reversal of
the otherwise expected "normal ordering" of Fig. 1.
The situation is different for $\Lambda(1405)$ - $\Lambda(1520)$
and
$\Lambda(1600)$, as the flavour state of  $\Lambda(1405)$ - $\Lambda(1520)$ is
totally antisymmetric. Because of this the
$\Lambda(1405)$ - $\Lambda(1520)$ gains an
attractive energy, which is
comparable to that of the $\Lambda(1600)$, and thus the ordering
suggested by the confining oscillator interaction is maintained.\\

If one goes beyond the schematic - but analytical - calculation
above, one should parameterize the short range parts of the
interaction (the long range parts are fixed by meson masses),
extract approximate meson-quark coupling constants from the known
meson-baryon ones and solve with computer 3 - body equations.
Such a program,  with a semirelativistic Hamiltonian (i.e.
kinetic energy is taken in a relativistic form)  and with the linear
confinement, has been realized in refs. \cite{GPVW,WGPV}. In the former 
case \cite{GPVW} only the spin-spin force of GBE interaction is
included, while in the latter one \cite{WGPV}
ps-, vector- and scalar-exchanges are considered with
  spin-spin, tensor 
and central force components. 
The spectra in both cases look pretty much the same,
which is achieved by a slight readjustment of the cut-off parameters
in the latter case, see Fig. 4.\\

It is clear that the higher Fock components $QQQ\pi, QQQK, ...$
(including meson continuum) cannot be completely integrated out in favor
of the meson-exchange $Q-Q$ potentials for some states above or near
the corresponding meson thresholds. Such  components, in addition to the
main one $QQQ$, could explain e.g. an exceptionally large splitting of the
flavour singlet states $\Lambda(1405)-\Lambda(1520)$, since the $\Lambda(1405)$
lies below the $\bar K N$ threshold and can be presented as  $\bar K N$ bound
system \cite{DAL}. Note, that in the case of the present approach this old
idea is completely natural and does not contradict a flavour-singlet  $QQQ$ 
nature of $\Lambda(1405)$ (it simply means that both QQQ and QQQK components
are significant in the present case) 
while it would be in conflict with naive constituent
quark model where no room for mesons in baryons. The alternative explanation
of the latter extraordinary big LS splitting would be that there
is some rather large spin-orbit force specific to the flavour-singlet
state only, which is also not ruled out.
 
An admixture of higher Fock components
will  be  important in order to understand strong decays of 
some excited states, especially in the case where the threshold
in the decay channel is close to the resonance energy.
While technically inclusion of such  components in addition to the main
one $QQQ$ in a coupled-channel approach is rather difficult task, it should
be considered as one of the most important future directions.

\begin{figure}[p]
\psfig{file=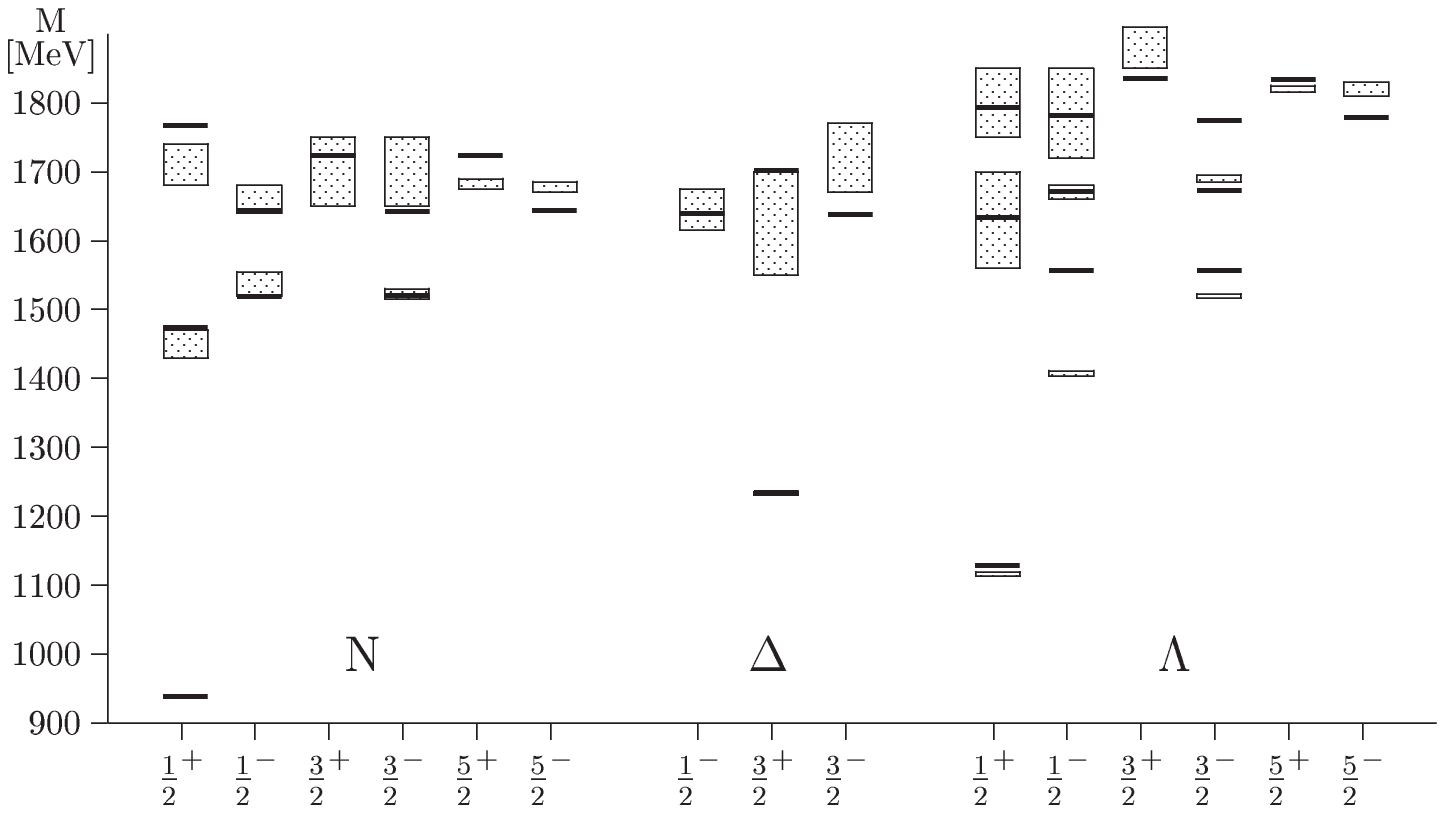,width=\textwidth}
\end{figure}
\begin{figure}[p]
\vspace*{-0.5cm}
\psfig{file=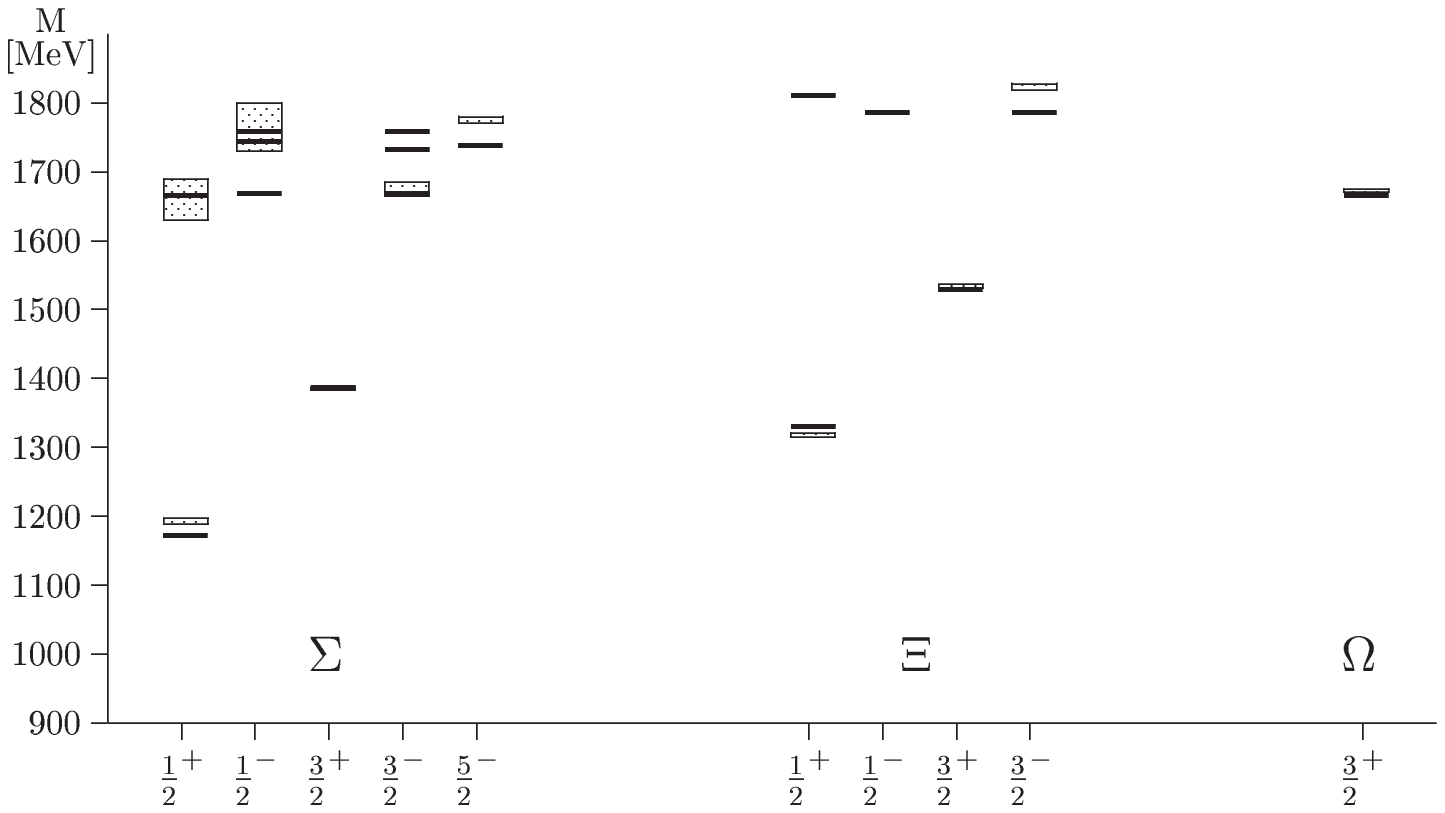,width=\textwidth}
\caption{Energy levels of the low-lying light- and strange-baryon
  states
with total angular momentum and parity $J^P$ (solid lines). The shadowed boxes
represent the experimental values with their uncertainties.}
\label{spectrum}
\end{figure}

\section{Instead of a conclusion}

Similar conclusions, that it is a GBE force which is
responsible for $\Delta - N$ splitting have been obtained
in a recent lattice study \cite{LIU}. A phenomenological
analysis of the L=1 negative parity spectra \cite{CG}
as well as $1/N_c$ expansion studies of L=1 nonstrange
spectra and of mixing angles obtained in strong and
electromagnetic decays \cite{CARONE}, also give a credibility to the
interaction (\ref{GBE}).

Finally, it is worth to mention, that this quark-quark interaction in the
baryon-baryon systems provides
a strong short-range repulsive core  \cite{STANCU,SHIM}.


\begin{thebibliography}{9}
\bibitem{SAH} Ya. B. Zeldovich and A. D. Sacharov, Yad. Fiz.
{\bf 4} (1966) 395 [Sov. J. Nucl. Phys. {\bf 4} (1967) 283];
A. D. Sacharov, JETP Lett. {\bf 21} (1975) 258.
\bibitem{RGG} A. DeRujula, H. Georgi and S.L. Glashow,
Phys. Rev. {\bf D12} (1975) 147.
\bibitem{MIT} A. Chodos, R. L. Jaffe, K. Johnson, C.B. Thorn,
V. Weisskopf, Phys. Rev. {\bf D9} (1974) 3471.
\bibitem{IK} N. Isgur and G. Karl, Phys. Rev. {\bf D18} (1978) 4187;
ibid. {\bf D19} (1979) 2653.
\bibitem{HOOFT} G. 't Hooft, Phys. Rev. {\bf D14} (1976) 3432.
\bibitem{KOCH} N. I. Kochelev, Sov. J. Nucl. Phys. {\bf 41} (1985) 291.
\bibitem{SR} E. V. Shuryak and J. L. Rosner, Phys. Lett. {\bf B218}
(1989) 72.
\bibitem{BLASK} W. H. Blask et al, Z. Phys. {\bf A337} (1990) 327.
\bibitem{AOKI} S. Aoki et al, Phys. Rev. Lett. {\bf 82} (1999) 4392.
\bibitem{WEIN} S. Weinberg, Physica {\bf 96A} (1979) 327.
\bibitem{MG} A. Manohar and H. Georgi, Nucl. Phys. {\bf B234} (1984) 189.
\bibitem{DP} D. I. Diakonov and V. Yu. Petrov, 
Nucl. Phys. {\bf B272} (1986) 457.
\bibitem{GR} L. Ya. Glozman and D. O. Riska,
Physics Reports {\bf 268} (1996) 263.
\bibitem{NJL} Y. Nambu and G. Jona-Lasinio, Phys. Rev. {\bf 122} (1961) 345;
ibid. {\bf 124} (1961) 246.
\bibitem{HOOFT2} G. 't Hooft, hep-th/9903189.
\bibitem{GV} L. Ya. Glozman and K. Varga, hep-ph/9901439, to appear
in Phys. Rev. D.
\bibitem{GLOZ} L.Ya. Glozman, Phys. Lett. {\bf B459} (1999) 589.
\bibitem{G} L.Ya. Glozman, Surveys in High Energy Physics -
in print, hep-ph/9805345. 
\bibitem{WGPV} R.F. Wagenbrunn, L.Ya. Glozman, W.Plessas, K. Varga,
these proceedings.
\bibitem{RB} D.O. Riska and G.E. Brown, hep-ph/9902319.
\bibitem{GPVW} L. Ya. Glozman,  W. Plessas, K. Varga, R. Wagenbrunn, 
Phys. Rev. {\bf D58} (1998) 094030.
\bibitem{DAL} R. H. Dalitz and A. Deloff, J. Phys. {\bf G 17} (1991) 289
(and references therein).
\bibitem{LIU} K. F. Liu et al, Phys. Rev {\bf D59} (1999) 112001.
\bibitem{CG} H. Collins and H. Georgi, Phys. Rev. {\bf D59}
(1999) 094010.
\bibitem{CARONE} C.E. Carlson et al, Phys. Rev. {\bf D59} (1999)
114008; C.D. Carone, hep-ph/9907412.
\bibitem{STANCU} Fl. Stancu, S. Pepin, and L. Ya. Glozman,
Phys. Rev. {\bf C56} (1997) 2779 [E: {\bf C59} (1999) 1219; 
Phys. Rev. {\bf D57} (1998) 4393; nucl-th/9906058.
\bibitem{SHIM} K. Shimizu and L.Ya. Glozman, nucl-th/9906008.
\end{thebibliography}
\end{document}